\documentclass[aps,floatfix,
prb,
reprint,
superscriptaddress
]{revtex4-2}

\usepackage{amsmath}
\usepackage{graphicx}
\usepackage{bm}
\usepackage{xcolor}
\usepackage[normalem]{ulem}
\DeclareUnicodeCharacter{2009}{ }
\usepackage{comment}

\usepackage{scrextend}

\usepackage{dcolumn}
\usepackage[caption=false, position=top]{subfig}

\newcommand{\eq}[1]{Eq.~(\ref{eq:#1})}
\newcommand{\fig}[1]{Fig.~\ref{fig:#1}}

\newcommand{\tab}[1]{Table~\ref{tab:#1}}

\newcommand{\vct}[1]{{\bf #1}}

\begin{document}

\title{Neural network-based nodal structures optimization
for interacting fermionic systems.}

 \author{William Freitas}
\affiliation{%
Gleb Wataghin Institute of Physics,
University of Campinas (UNICAMP), 13083-970 Campinas, SP, Brazil}

\author{B.~Abreu}
\affiliation{Pittsburgh Supercomputing Center, Carnegie Mellon University, Pittsburgh, PA 15213, USA}

\author{S.~A.~Vitiello}
\affiliation{%
Gleb Wataghin Institute of Physics,
University of Campinas (UNICAMP), 13083-970 Campinas, SP, Brazil}
   
\begin{abstract}

Simulating strongly correlated fermionic systems remains a fundamental challenge in quantum physics, largely due to the sign problem in quantum Monte Carlo (QMC) methods. We present a neural network-based variational Monte Carlo (NN-VMC) approach, leveraging a flexible neural network ansatz to represent the many-body wavefunction. Focusing on quantum dots with up to 30 electrons, we demonstrate that NN-VMC significantly reduces variational bias and achieves ground-state energies surpassing those of fixed-node diffusion Monte Carlo (DMC). A key feature is that the neural network adaptively learns and optimizes nodal structures during energy minimization. We provide qualitative insights into the nodal structure of fermionic wavefunctions by comparing the nodal structures generated by NN-VMC with those obtained from traditional trial functions. Additionally, we reveal spin-resolved radial distributions and electron density profiles, highlighting the versatility and accuracy of NN-VMC. This work underscores the potential of machine learning to advance quantum simulations and deepen our understanding of strongly correlated systems.
\end{abstract}

\maketitle

\section{Introduction}  
Accurately modeling strongly correlated fermionic systems remains a central challenge in quantum many-body physics, with implications for systems ranging from quantum chromodynamics and nuclear matter to ultracold gases and condensed matter systems \cite{pas20,adl24,bay23,car21,blo22}. These systems exhibit strong correlations that traditional methods struggle to capture, requiring many-body correlations in the trial function. Quantum Monte Carlo (QMC) methods \cite{wag16} have proven particularly effective, as they leverage non-relativistic quantum mechanics to capture intricate many-body effects.

Quantum dots (QDs) are versatile platforms for studying strong correlations and excitonic states \cite{sik89,mak90,hsi24}, with applications in scalable quantum computing and other quantum technologies \cite{esp24,joh24,mey23,bor23,bor23b,law23}. Furthermore, QDs provide a unique platform for studying generalized Fermi-Hubbard physics, offering insights into condensed matter phenomena and advancing our understanding of quantum many-body systems \cite{wan22,kic22}.

Machine learning (ML) offers a powerful alternative to traditional trial functions in QMC methods by enabling more flexible and expressive wavefunction representations. Deep neural networks (DNNs) capture complex many-body correlations more effectively than conventional approaches, enhancing wavefunction accuracy and reducing variational bias \cite{joh22,zhe23,lou24}. This adaptability makes ML-based methods particularly valuable for refining nodal structures and mitigating the limitations of fixed-node approximations, which constrain random walks to regions where a predefined wavefunction retains a single sign. By learning more accurate nodal surfaces, ML approaches achieve better approximations of the true wavefunction \cite{pfa20,fu24}, ultimately improving the accuracy of QMC simulations.

The fixed-node approximation mitigates the sign problem but introduces an intrinsic bias due to the predetermined nodal structure \cite{car17,cos98,yu17,ras15,naz16}. Its accuracy depends heavily on the quality of the trial wavefunction´s nodal surfaces, as highlighted by Ceperley \cite{cep91}. The computational complexity of the sign problem has been widely studied, with Troyer and Wiese \cite{tro05} analyzing its fundamental limitations. Additionally, Foulkes \textsl{et al.} \cite{fou01} examined its impact on solid-state simulations, providing further context for its challenges. Moreover, traditional trial wavefunctions impose rigid nodal surfaces, limiting their flexibility and accuracy. This is particularly evident in regions near the nodal hypersurfaces, where traditional trial functions often exhibit significant inaccuracies.

Building on these insights, Carleo and Troyer \cite{car17sc}
revolutionized the field by introducing neural network quantum states
(NNQS), demonstrating their capacity to accurately represent many-body
wavefunctions and solve quantum spin systems. This breakthrough paved
the way for applications such as deep neural network solutions for the
electronic Schr\"odinger equation \cite{her20} and the extension \cite{pfa20} of these methods to ab initio simulations of many-electron systems. Alternative architectures, notably recurrent neural networks, have been proposed and demonstrated to provide advantages, since their autoregressive nature facilitates the efficient computation of physical estimators \cite{car20}. Furthermore, Luo and Clark \cite{luo19} demonstrated the power of neural networks in refining nodal structures through backflow transformations, while Choo et al. \cite{cho18prl} explored how NNQS can capture symmetries and excitations in many-body systems. These contributions highlight the transformative potential of ML-based approaches in overcoming the limitations of traditional QMC methods, particularly in addressing the sign problem and enhancing the accuracy of quantum many-body simulations.

In this work, we employ neural network-based variational Monte Carlo (NN-VMC) \cite{pfa20} to implicitly optimize nodal structures during energy minimization. Focusing on quantum dot systems with up to 30 electrons, we demonstrate that NN-VMC reduces variational bias and achieves ground-state energy estimates that surpass those obtained with diffusion Monte Carlo (DMC) using the fixed-node approximation. More importantly, we provide qualitative insights into the nodal structure of fermionic wavefunctions. Unlike previous works on molecular systems, which did not compare the nodal structures generated by NN-VMC and traditional trial functions \cite{fu24,ren23}, our study provides such a comparison. Furthermore, we provide new insights into strongly correlated quantum dot systems, such as spin-resolved radial distribution functions and electron density profiles, highlighting the broader applicability of ML-based wavefunction representations in quantum many-body physics. Our findings establish the broader applicability and effectiveness of NN-VMC methods for studying strongly correlated fermionic systems, marking a significant step forward in the field. Moreover, these NN representations can be efficiently computed using graphical processing units (GPUs), and their derivatives can be calculated without finite difference errors using automatic differentiation.

Given the challenges associated with QMC methods and the crucial role of nodal structure optimization, we adopt a neural network-based variational Monte Carlo (NN-VMC) approach. By leveraging the flexibility of deep neural networks, we refine wavefunction representations to improve accuracy in strongly correlated fermionic systems. In the next section, we detail the NN-VMC framework, including the modeling of trial wavefunctions, optimization strategies, and key computational techniques applied to quantum dot systems.

\section{Methods} Our NN-VMC method models trial wavefunctions using a neural network ansatz, optimizing them based on variational principles and computing physical properties through Monte Carlo sampling. Below, we describe the system model, Hamiltonian, neural network architecture, and the optimization process that enables adaptive nodal structure refinement.

We investigate a system of circular QDs made from semiconductor heterostructures, which is modeled by electrons interacting via the Coulomb potential and confined within a two-dimensional harmonic trap. With $N$ electrons placed in the $z=0$ plane, this model's Hamiltonian is 

\begin{equation}\label{eq:hamil}
\mathcal H = 
\frac {-\hbar^2}{2m^*} \sum^N\limits_{p=1} \bm\nabla_p^2 + 
\frac {m^* \omega^2}{2} \sum^{N}\limits_{p=1}r_{p}^2 +
\frac {k_e e^2}{\varepsilon} \sum^N\limits_{p<q} \frac {1}{r_{pq}} \ ,
\end{equation}

\noindent where $m^*$ is the effective mass the electron, $\hbar$ is the reduced Planck constant, $\omega$ is the angular frequency of the harmonic trap, $\varepsilon$ is the dielectric constant of the semiconductor, $e$ is the electron charge. The indices $p$ and $q$ identify two spatial points corresponding to the $N$ electrons in the system. The distance between these points is given by $r_{pq} =\lvert\mathbf r_p - \mathbf r_q\rvert$ where $(p,q) \in [1,2, ..., N]$.
It is convenient to define $R$ as the set of all such points. A dimensionless Hamiltonian form $\hat{\mathcal H}$ used in this work is obtained by the introduction of an effective Hartree energy scale $E_{\rm H}^* = \hbar^2/ m^* (a_{\rm B}^*)^2$, where $a_{\rm B}^*$ is the effective Bohr radius, $a_{\rm B}^* = \varepsilon\hbar^2/ k_e m^* e^2$ and $k_e$ is the Coulomb constant. Therefore, the effective Bohr radius and Hartree energy allow us to express the Hamiltonian in a dimensionless form. This transformation simplifies parameter dependencies and enables efficient energy optimization within the NN-VMC framework:

\begin{equation}\label{eq:redhamil}
\hat{\mathcal H} = 
-\frac {1}{2} \sum^N\limits_p \hat{\bm\nabla}_{\hat{\vct{ r}}_p}^2 + 
\frac {1}{2\lambda^2} \sum^{N}\limits_{p}\hat{r}_{p}^2 +
\sum^N\limits_{p<q} \frac {1}{\hat{r}_{pq}} \ ,
\end{equation}

\noindent where $\hat{r}_{p} = r_{p}/a_{\rm B}^*$ and $\lambda=E_{\rm H}^*/\hbar\omega$ which dictate the relative importance of the confining potential and the Coulomb interaction. For simplicity, we omit the hat notation for spatial coordinates from now on.

We choose parameters based on a quantum dot of gallium arsenide, GaAs, which has an electronic effective mass $m^* = 0.067m_e$ ($m_e$ is the electron mass), and a dielectric constant $\varepsilon = 12.7$ \cite{har05}. For the confining trap, we chose $\hbar\omega = 0.28 E_{\rm H}^*$, a typical value in the literature investigating the behavior of quantum dots \cite{ped00}.

To represent the wave function of this fermionic system, we employ a neural network architecture based on the FermiNet approach \cite{spe20,fre23q,fre24}, which provides a highly flexible representation of complex wave functions. Despite requiring extensive training, this approach offers superior representational power compared to other leading neural network architectures for fermionic systems \cite{her20, luo19}. After optimization, the FermiNet-based trial function was then used to perform estimations of the observables of interest, such as the total energy. The trial wave function is given by

\begin{align}
\psi(R)
=
\sum\limits_k & \omega_k 
\det 
\left[
\phi_{ij}^{k\uparrow} (R) \right]
\times
\det
\left[
\phi_{ij}^{k\downarrow} (R)
\right]\ ,
\label{eq:psi}
\end{align}    
where $\omega_k$ are variational parameters and orbitals $\phi_{ij}^{k\alpha}$ are constructed using artificial NNs. The indices $i$ and $j$ identify the rows and columns of the matrix over which the determinant is being computed,
where $(i,j) \in [1,2,\ldots,N/2]$. Moreover, the orbitals are given by

\begin{align}
\phi_{ij}^{k\alpha} &
(R)
= (\mathbf w_{i}^{k} \cdot \mathbf h_{j}^L + g_{i}^k ) \pi_{i}^k \exp(-|\bm\Lambda_{i}^{k} \mathbf r_{j}|) \ ,
\label{eq:orb}
\end{align}
where $i$ is not associated with any spin, the spin $\alpha$ of index $j$ is the same as in $\phi_{ij}^{k\alpha}$, $\mathbf h_{j}^{L}$ are neural network outputs and they are associated with the spatial point $j$ that corresponds to the same spin $\alpha$ of $\phi_{ij}^{k\alpha}$, while $\mathbf w_{n}^{k}, g_{n}^{k}, \pi_{n}^{k},$ $\bm\Lambda_{n}^{k}$ are variational parameters.

Hence, the trial wave function is parameterized using a neural network ansatz following the FermiNet architecture, as given in Eqs. (\ref{eq:psi})–(\ref{eq:orb}). The input features, layer propagation equations, and interaction terms that define the neural network outputs are detailed in the appendix \ref{ap:trial-ann}. These include the transformation of electron coordinates, activation functions, and network layer updates necessary for efficient energy optimization.
The Kronecker-factored approximate curvature (KFAC) algorithm \cite{mar15} is used to optimize the variational parameters of the trial wave function, ensuring stable convergence.

\begin{figure}[ht]  
    \centering
    \includegraphics[width=0.45\textwidth]{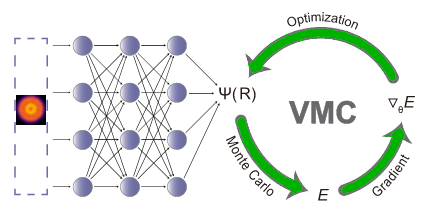}
    \caption{Schematic illustration of the NN-VMC method. Adapted from Ref.~\cite{fu24}. }
    \label{fig:algorithm}
\end{figure}

\fig{algorithm} illustrates the neural network-based variational Monte Carlo (NN-VMC) method applied to quantum dot systems. In this approach, the spatial coordinates of electrons, confined by a harmonic potential, are used as inputs to a neural network to represent the trial wavefunction $\psi(R)$. The algorithm iteratively optimizes the variational parameters of the trial function to minimize the expected energy $\langle E \rangle$ using gradient-based methods.

The Metropolis algorithm is employed to generate new electron configurations, which are subsequently fed back into the optimization loop. This iterative process continues until convergence is achieved, ensuring that the trial function approaches the true ground state with improved accuracy.
This methodology leverages the expressiveness of neural networks to refine nodal structures and capture complex quantum many-body interactions.

It is important to note that for 
$|\mathbf{r}_i| \rightarrow 0$, the harmonic confinement potential used in this study does not introduce essential singularities. 
The inclusion of $\mathbf r_i$
in the ansatz is designed to enhance the flexibility of the neural network model to adapt to different spatial configurations rather than to address any singularity in the trap potential. In two and three dimensions, the kinetic energy operator inherently includes terms proportional to $1/r^2$, which naturally regulate divergences at small $r$. An additional $1/r$ dependence in the kinetic energy arising from the trial function does not introduce unbounded states in this case. The neural network-based approach ensures smooth approximations of the wavefunction through activation functions and optimization mechanisms that minimize energy while satisfying the variational principle.

Any potential discontinuity at $|\mathbf r_i| \rightarrow 0$ is suppressed during training, as contributions of non-physical kinetic energy are penalized in the optimization process. The variational parameters, including those that influence terms involving $|\mathbf r_i |$, are adjusted to minimize energy, ensuring that the trial function remains physically valid. The resulting energies and statistical uncertainties confirm that the ansatz does not lead to unbounded kinetic energy contributions. However, further analysis of the behavior of $|\mathbf r_i |$ terms near $|\mathbf r_i| =0$ could provide deeper theoretical insights into their impact on wavefunction optimization.
 
This methodology enables the adaptive optimization of nodal structures and the accurate computation of ground state energies, as demonstrated in the results section.

\section{Results} We investigated the effectiveness of using machine learning, specifically neural networks, in representing wave functions for fermionic system. Our focus was on QDs with up to 30 electrons, and we compared the results of our neural network-based variational Monte Carlo (NN-VMC) approach with the diffusion Monte Carlo (DMC) method using the fixed-node approximation.

\begin{figure}[ht]
    \centering
\includegraphics[width=0.4\textwidth]{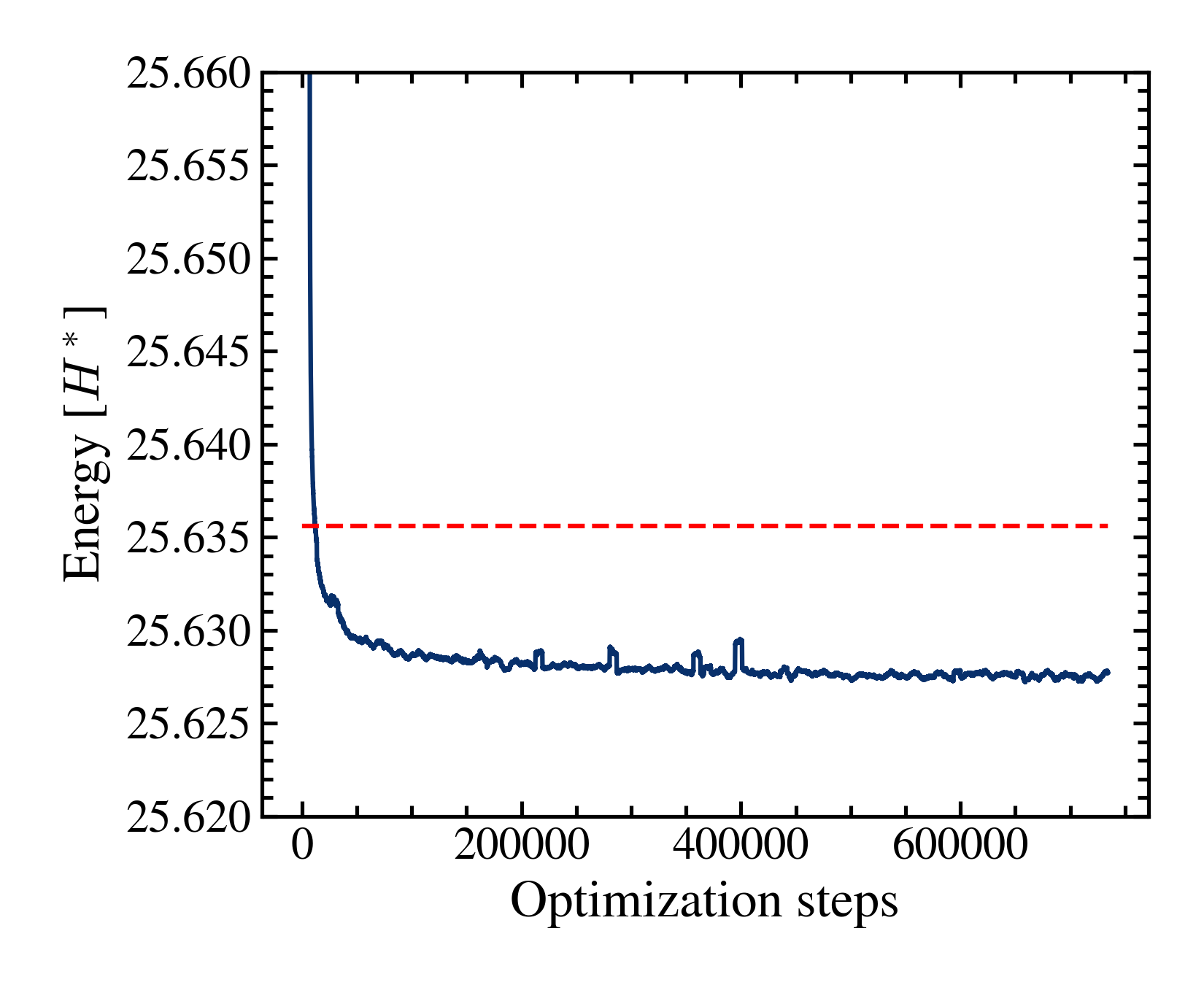}
    \caption{Evolution of the variational total energy of a QD with 12 electrons in the NN-VMC algorithm. The dashed red line stands for a DMC result using the fixed-node approximation \cite{loh11}. The spikes in the variational total energy occur when the system escapes a local minimum.}
    \label{fig:opt-ene}
\end{figure}

The evolution of the total variational energy for a 12-electron QD is shown in \fig{opt-ene}. The energy decreases rapidly initially, followed by a slower decline, reaching values below DMC results after about 40,000 optimization steps. Further refinement requires hundreds of thousands of steps, highlighting the computational cost of high accuracy. As the electron count increases, the computational burden grows significantly, making calculations for systems with more than 30 electrons demanding. Nevertheless, NN-VMC achieves a statistically significant improvement (0.5\%) over DMC, with smaller energy uncertainties. This improvement stems from NN-VMC's ability to adaptively refine nodal surfaces, capturing complex correlations that traditional methods struggle to represent. 

Moreover, compared to PIMC, NN-VMC also reduces kinetic energy contributions. Kinetic energies obtained using the NN representation of the trial function can be compared with those from a PIMC simulation in \tab{ke}. As shown, the kinetic energies obtained with ML are systematically lower than those estimated through the PIMC method. For completeness, \tab{te_ho} presents the total energies for QDs with electrons described by the trial function of the harmonic oscillator.

\begin{table}[ht]
\caption{Kinetic energies for QDs with $N$ electrons in units of $E_H^*$. Each column presents results for the corresponding number of electrons. The second row displays results obtained using the NN representation of the trial function, while the last row shows results for the harmonic oscillator, both obtained in this work. The third row presents results from a PIMC simulation.}
\label{tab:ke}
\begin{ruledtabular}
\begin{tabular}{lcccc}
              & $N=6$          & $N=12$  & $N=20$         & $N=30$          \\  
$K_{\rm NN}$   & 0.9406(3)  & 2.2238(4)   & 4.2250(6)  & 7.061(1)    \\ 
$K_{\rm PIMC}\footnote{\cite{kyl017}}$ & 0.94071(9) & 2.2402(6)   & 4.304(2)   & -           \\ 
$K_{\rm HO}$   & 0.6614(1)  & 1.6266(2)   & 3.126(2)   & 5.476(6)    \\
\end{tabular}
\end{ruledtabular}
\end{table}

\begin{table}[ht]
\caption{Total energies for QDs with $N$ electrons described by trial functions of the harmonic oscillator in units of $E_H^*$. Each column presents results for the given number of electrons. }
\label{tab:te_ho}
\begin{ruledtabular}
\begin{tabular}{lcccc}
              & $N=6$          & $N=12$  & $N=20$         & $N=30$          \\ 
$E_{\rm HO}$   & 8.070(1)   & 26.670(1)   & 63.817(6)  & 127.14(3) 
\end{tabular}
    \end{ruledtabular}
\end{table}

\begin{figure}[ht]  
    \centering
\includegraphics[width=0.9\linewidth]{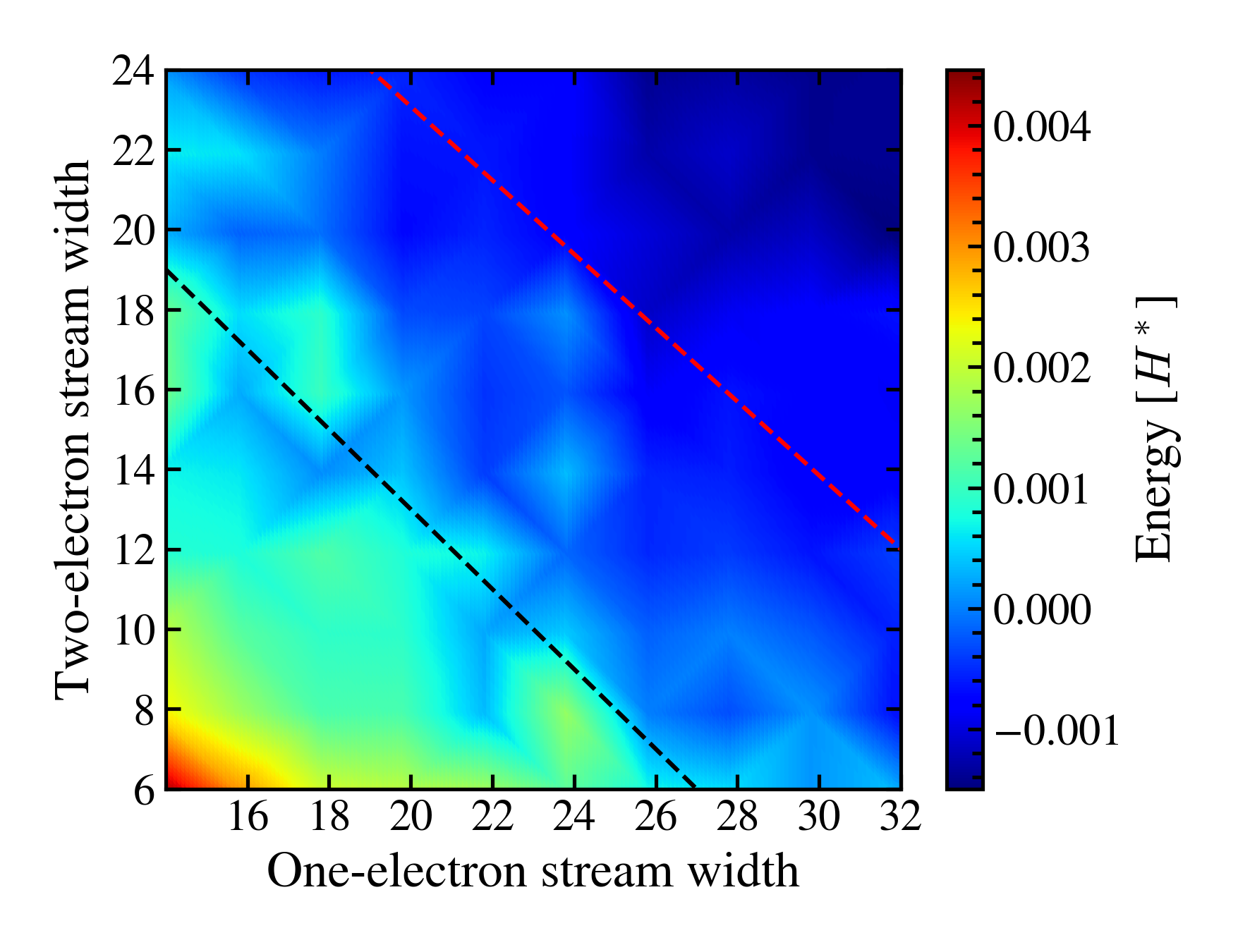} \caption{
 Color map of the total energy as a function of the widths of the one-electron and two-electron streams. The dashed lines divide the figure into three distinct regions. The dark blue region corresponds to hyperparameter values that yield results surpassing those obtained by the DMC method. The lower region, predominantly green, represents hyperparameter values that produce results inferior to those of the DMC method. The intermediate region shows results where the NN-VMC method begins to achieve energies nearly identical to those obtained using the DMC method.}
\label{fig:map-energy}
\end{figure}

The hyperparameters were carefully selected based on empirical tuning.
The architecture is characterized by four primary hyperparameters: The total number of layers $L$, which determines the depth of the network.
The widths of the one- and two-electron streams.
The number of multi-orbital expansions, which corresponds to the maximum $k$ in Eq. \ref{eq:psi}.
For systems with $N\geq12$ electrons, typical hyperparameter values are as follows:
4 neural network layers,
a width of 128 for the one-electron stream,
a width of 32 for the two-electron stream, and
8 determinants in the multi-orbital expansion.
These hyperparameter values directly impact the accuracy of the trial wavefunction representation. \fig{map-energy} presents results for a system of 6 electrons, illustrating how the energy difference between DMC and our results depends on the widths of the one-electron and two-electron streams.

After the optimization process, we select the optimized set of variational parameters and perform a standard variational Monte Carlo integration to further reduce statistical uncertainties. 
\tab{ene} compares the total energies obtained by NN-VMC with those obtained by the leading methods used in the investigation of QDs, DMC, path-integral Monte Carlo (PIMC), coupled cluster with singles and doubles (CCSD) and in-medium similarity renormalization group (IM-SRG). Particularly, the CCSD and IM-SRG methods do not provide an upper bound for the total energy of the system. Across all electron counts, our approach consistently yields lower energy values, except when comparing with IM-SRG for a few electrons. For the 30-electron system, the total energy calculated using our neural network approach is 123.9469(2) $E_{\rm H}^*$, compared to 123.9683(2)  $E_{\rm H}^*$ from DMC, representing an improvement of 0.02  $E_{\rm H}^*$. Our results highlight the consistent improvements achieved by the NN-VMC method.

\begin{table}[ht]
    \caption{
Total energies for QDs with $N$ electrons in units of $E_{\rm H}^*$. Each column shows results for the given number of electrons, the second row presents results obtained in this work. The third, fourth, fifth, and sixth rows display results from the literature for the given method.
    }
\begin{ruledtabular}

\begin{tabular}{lcccc} 
& $N=6$ & $N=12$ & $N=20$ & $N=30$ \\
This work & 7.59703(9) & 25.62599(6) & 61.9073(1) & 123.9372(1) \\
{DMC\footnote{\cite{loh11}, $^*$\cite{hog13};\quad \textsuperscript{b}\cite{kyl017}; \quad\textsuperscript{c}\cite{yua17} }}  & 7.6001(1) & 25.6356(1) & 61.922(2) & 123.9683(2)$^*$ \\
PIMC$^b$ & 7.5980(1) & 25.6456(6) & 61.992(3) & {\rm N/A} \\
CCSD$^c$ & 7.6341 & 25.7345 & 62.1312 & 124.3630 \\
IM-SRG$^c$ & 7.5731 & 25.6259 & 61.9585 & 124.1041\\ 
    \end{tabular}
    \end{ruledtabular}
    \label{tab:ene}
\end{table}

To evaluate the nodal structures generated by our neural network representation, we compared the wave function with a harmonic oscillator trial function, the usual Slater determinant of non-interacting harmonic oscillator orbitals with the harmonic frequency as a variational parameter. The primary objective of this study is to demonstrate the effectiveness of NN-VMC in reducing variational bias and achieving lower ground state energies compared to traditional methods like DMC. The energy minimization process is the central metric for evaluating the success of your approach. A detailed quantitative analysis of the nodal structure, while valuable, is not directly necessary to achieve this primary goal. Moreover, the sophisticated analytical tools necessary for such quantitative analysis  would significantly extend the scope of the study. The qualitative comparison of nodal structures provided in the study offers sufficient initial validation of the NN-VMC method's ability to produce smoother and more symmetric nodal surfaces. These insights highlight the adaptability of neural networks in reducing variational bias.

\begin{figure*}[ht]
    \centering
    \includegraphics[width=0.9\linewidth]{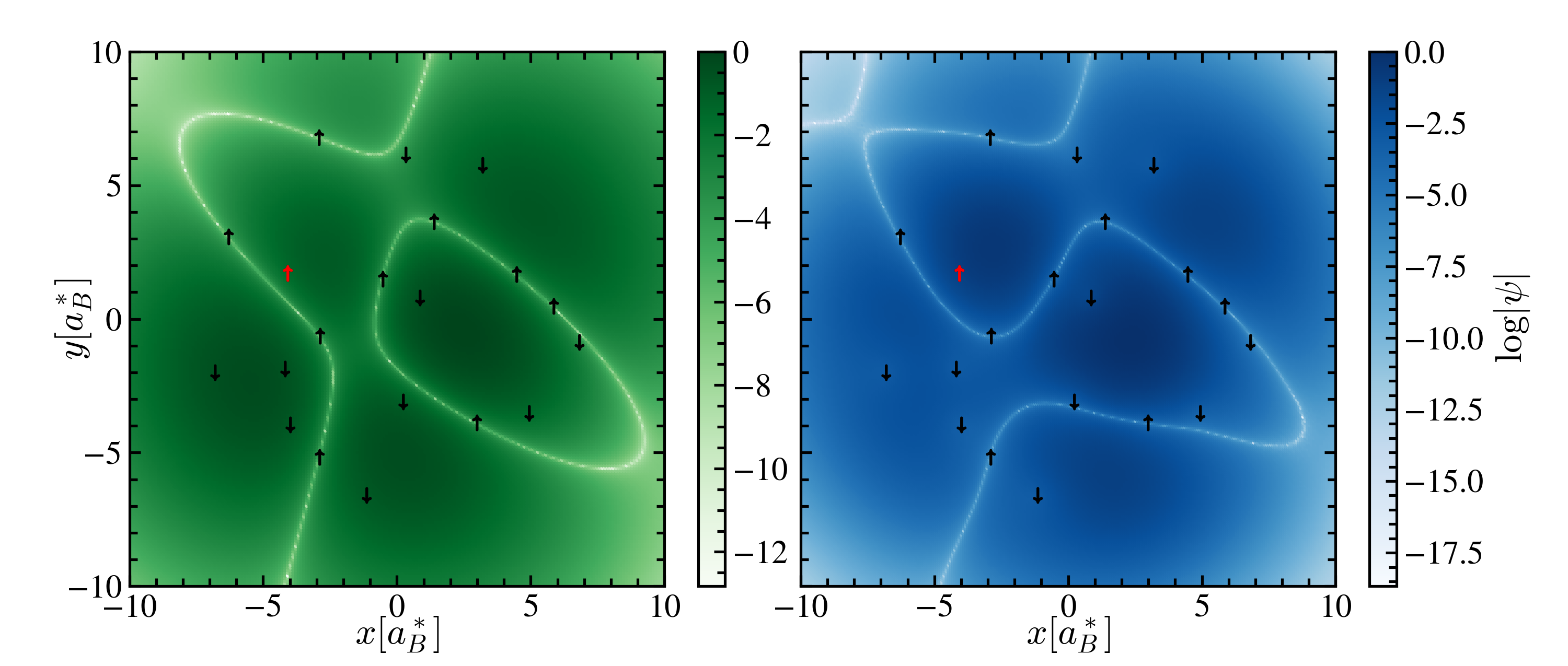}
\caption{Nodal structure of trial functions of a QD with $N=20$ electrons, with the white lines indicating the nodal hypersurfaces. (Left) A trial function of the harmonic oscillator form. (Right) Neural network representation of the ML-learned wave function. Up (down) arrows stand for up (down) spins. The red arrow depicts the electron that is moved over the hyperplane, all others being fixed. }
    \label{fig:node-struct}  
  \end{figure*}

The nodal structure is not explicitly prescribed but is implicitly learned by the neural network through energy minimization. Defining a generalizable quantitative curvature metric is non-trivial and remains an open area of research \cite{ren23}. As the network adjusts its parameters during training, it optimizes the placement and smoothness of the nodal surfaces to reduce the variational energy, leading to a self-consistent nodal topology that reflects the underlying physics of the system, \fig{node-struct} illustrates the smooth and symmetric nodal surfaces generated by NN-VMC. In contrast, traditional trial functions impose rigid nodal structures that lead to higher energy estimates (see the appendix \ref{ap:nodal-structures}).

The nodal regions of the neural network-derived hyperplanes in blue (right) are smoother and exhibit less curvature compared to the traditional trial function represented in green (left), leading to a reduction in kinetic energy and an overall more accurate representation of the system. The comparison was repeated by choosing different initial electronic configurations, also including electrons with spin down. The result was always consistent with the one we show in \fig{node-struct}. Naturally, changing the electron being moved impacts the overall aspect of the corresponding nodal structure, as can be seen in the appendix \ref{ap:nodal-structures}. The NN representation of the trial function shows the requirements of a ``good'' nodal structure, \textsl{i.e.}~it exhibits smoothness and symmetry. Traditional trial wavefunctions often impose rigid nodal surfaces, which can lead to high kinetic energy contributions and suboptimal energy estimates. In contrast, the neural network's flexibility allows it to adapt the nodal surfaces smoothly and symmetrically, as qualitatively depicted in \fig{node-struct}.

\begin{figure}[htbp]
    \centering
    \includegraphics[width=0.45\textwidth]{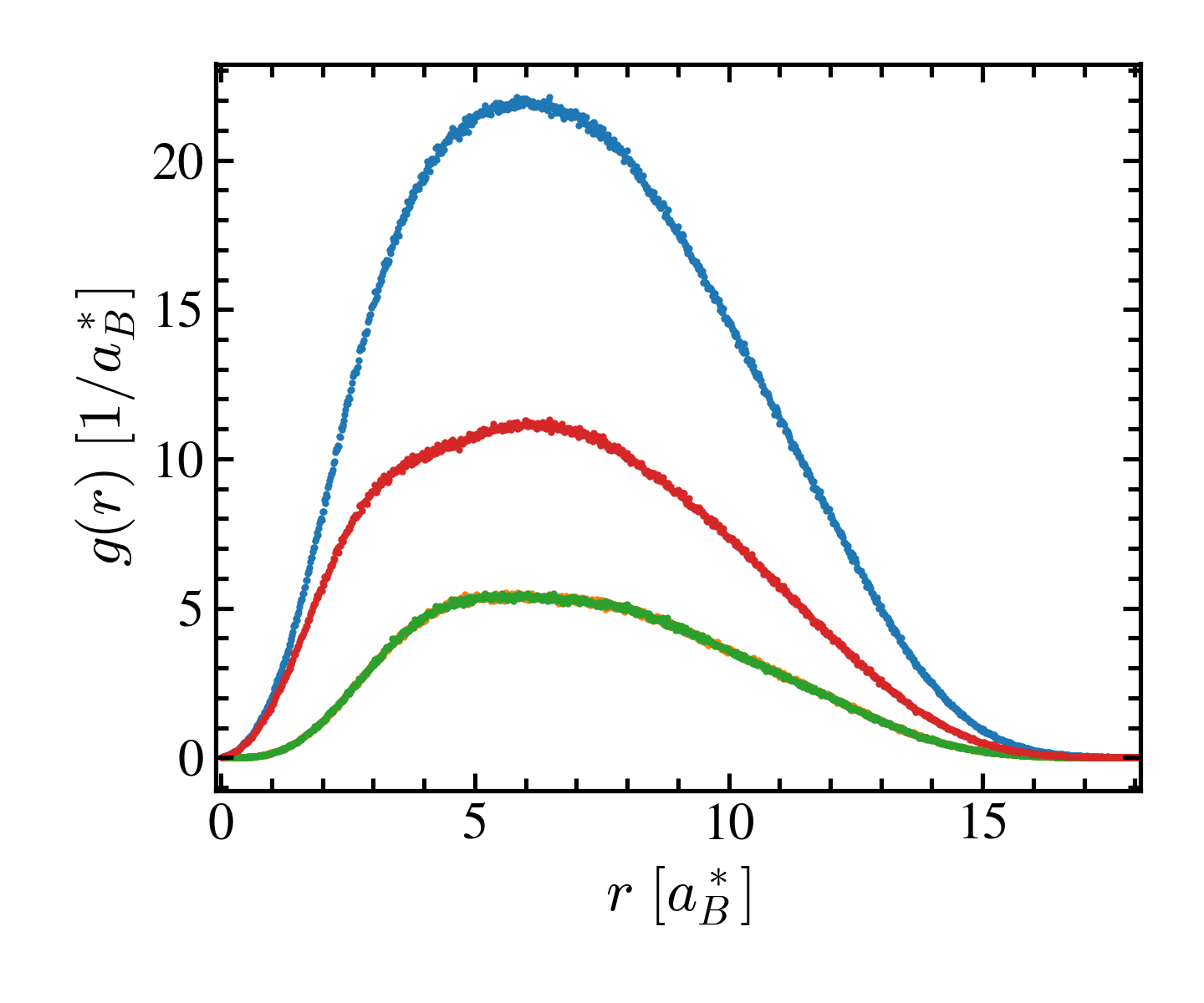}
    \caption{Pair distribution functions in units of $1/a_0^*$, the inverse of the effective Bohr radius, as a function of the electrons separation. The pair distribution functions for up-up spins and down-down spins are collapsed in the figure scale, they are visible in green. In red is depicted $g_{\uparrow \downarrow}(r)$ and in blue $g(r)$ of the electrons regardless of their spin.}
    \label{fig:pair-distr-2d}
\end{figure}

\begin{figure}[ht]
    \centering
    \includegraphics[width=0.475\textwidth]{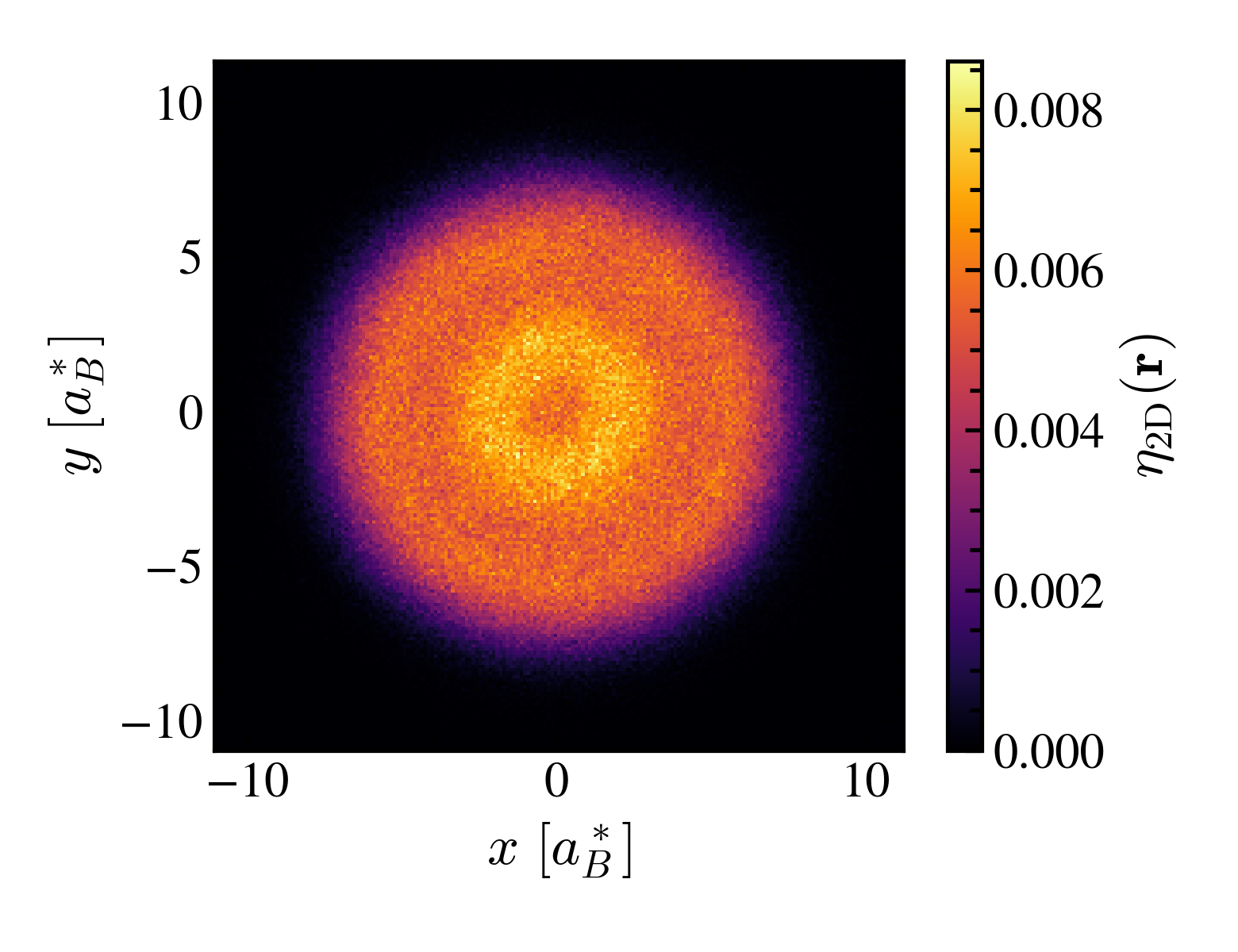}
    \caption{A normalized bi-dimensional histogram of the electronic positions of a QD of 20 electrons.}
    \label{fig:dens-prof-2d}
\end{figure}

Showcasing the ability of the neural-network wave function to yield accurate results for physical properties of the system, in \fig{pair-distr-2d} we show results for radial distribution functions of different relative electronic spin orientations. Furthermore, the two dimensional density profile for a QD can be computed through
\begin{align}
    \eta_{\rm 2D}(\vct r) = \sum_i \delta(\vct r_i - \vct r)\ ,
\end{align}
normalized by imposing $\int \eta_{\rm 2D}(\vct r) dA = 1$. The result for a QD with 20 electrons is displayed in \fig{dens-prof-2d}. Although circular symmetry was not explicitly imposed, our neural network was able to capture this feature during training. This reinforces the flexibility and effectiveness of our approach in representing complex wave functions without prior assumptions about the geometry of the system.  Figs.~1-4 provide new insights into the interplay between confinement potentials and electron correlations, showcasing the versatility of NN-VMC.

The small energy uncertainties (e.g., 0.00009 $E_H^*$ for a 12-electron system) highlight the robustness of the NN-VMC method and confirm the statistical significance of its improvements over DMC. These results confirm that our machine learning-based approach can reliably reduce variational bias in quantum Monte Carlo simulations.

\section{Conclusion} 
Several sophisticated approaches have been proposed to mitigate or circumvent the sign problem in QMC simulations of Fermi systems \cite{uly20,kor20,lev21,wei16,hof22}. The fixed-node approximation is a popular approach and has yielded good results for various systems \cite{and24,pes24,cur23,she22,rhi21,fra19,dit19,loh11,hog13}, but introduces an intrinsic bias through an {\sl a priori} chosen nodal structure. 

This work establishes NN-VMC as
a powerful tool for quantum simulation. It significantly improves upon
fixed-node DMC by directly optimizing nodal
structures, thereby reducing variational bias. Furthermore, NN-VMC
offers key advantages for calculating properties that do not commute
with the Hamiltonian; unlike DMC, it avoids the need for extrapolated
estimators, which are a significant source of bias. Our method is also
free from the population control bias introduced by fluctuating walker
numbers in DMC. We demonstrate these advantages by achieving
statistically superior ground-state energy estimates for quantum dots
with up to 30 electrons, highlighting the potential of machine learning
to overcome central challenges in many-body physics.

Future research will focus on optimizing the computational efficiency of NN-VMC to scale to larger systems and further refine the learned nodal structures. This includes investigating nodal curvature metrics, improving algorithmic scalability, and leveraging GPU-accelerated frameworks to enhance performance. Additionally, NN-VMC can be extended to study spectral properties and excited states \cite{pfa24}, broadening its applicability to strongly correlated fermionic systems.

Integrating machine learning with quantum simulations advances next-generation computational techniques, benefiting quantum computing and materials discovery. 

\textsl{Acknowledgements} - SAV and WF acknowledge financial support from the Brazilian agency, Funda\c{c}\~ao de Amparo \`a Pesquisa do Estado de S\~ao Paulo grants \#06837-1, \#2023/07225-0 and \#2020/10505-6, São Paulo Research Foundation (FAPESP). This work used Delta at the National Center for Supercomputing Applications through allocation CIS230072 from the Advanced Cyberinfrastructure Coordination Ecosystem: Services {\&} Support (ACCESS) program, which is supported by U.S. National Science Foundation grants \#2138259, \#2138286, \#2138307, \#2137603, and \#2138296 \cite{boe23}.

\appendix

\section{Detailed description of the\\
neural network architecture}\label{ap:trial-ann}

This section provides a detailed description of the neural network architecture used in the NN-VMC ansatz, including the input representations \eq{in}, the layer update rules  \eq{infoflow}, and the interaction terms, \eq{vinterm}. These equations govern how information propagates through the network to refine nodal structures during energy minimization.

 The two streams are constructed to manage the data flow: the one-electron stream $\vct h_{i}^{\ell}$, which processes inputs related to single-electron configurations, and the two-electron stream $\vct h_{ij}^{\ell}$, which handles inputs related to two-electron configurations; $\ell$ stands for the particular layer being considered. Both streams are characterized by their respective widths, referring to the number of units in each layer. 

The input for the first layer of each stream considers information from the electron positions and their relative coordinates

\begin{equation}
\label{eq:in}
\vct h_{i}^{0} = 
\left( \vct r_{i}, |\vct r_{i}| \right) \ ; \
\vct h_{ij}^{0} = 
\left( \vct r_i - \vct r_j , | \vct r_i - \vct r_j | \right) \ .
\end{equation}
The spin of the $j$-th electron may be the same as or different from that of the $i$-th electron.

Information in each stream propagates from layer 
 $\ell$ to $\ell+1$ according to the following update rules:

\begin{subequations}
\label{eq:infoflow}
\begin{align}
\vct h_{i}^{\ell+1} &= 
\tanh \left(
\vct V^\ell \vct f_{i}^{\ell} + \vct b^\ell
\right) + \vct h^{\ell}_{i} \ , \\
\vct h_{ij}^{\ell+1} &= 
\tanh \left(
\vct W^\ell \vct h_{ij}^{\ell} + \vct c^\ell
\right) + \vct h^{\ell}_{ij} \ ,
\end{align}
\end{subequations}
where $\vct V^\ell$ and $\vct W^\ell$ are the weight matrices, while $\vct b^\ell$ and $\vct c^\ell$ are biases and $\vct f_{i}^{\ell}$ represents intermediate vectors, defined as:

\begin{align}
\label{eq:vinterm}
\vct f^{\ell}_{i} = 
&\left(
\vct h_{i}^{\ell} ,
\frac 2 {N} \sum\limits_{m=1}^{N/2} 
\vct h_{m}^{\ell},
\frac 2 {N} \sum\limits_{j=1}^{N/2} 
\vct h_{j}^{\ell},\right.
\left.
\frac 2 N \sum\limits_{m=1}^{N/2} 
\vct h_{i m}^{\ell} ,
\frac 2 {N} \sum\limits_{j=1}^{N/2}
\vct h_{i j}^{\ell}
\right)
\ ,
\end{align}
where the summation over index $j$ is performed for electrons with spins opposite to those of index $i$, the summation over index $m$ is performed for particles with the same spin as index $i$.

\section{Nodal structure of the traditional trial function and the one using the NN representation}\label{ap:nodal-structures}

\begin{figure*}[ht]
\centering
\subfloat[]{\includegraphics[width=0.5\linewidth]{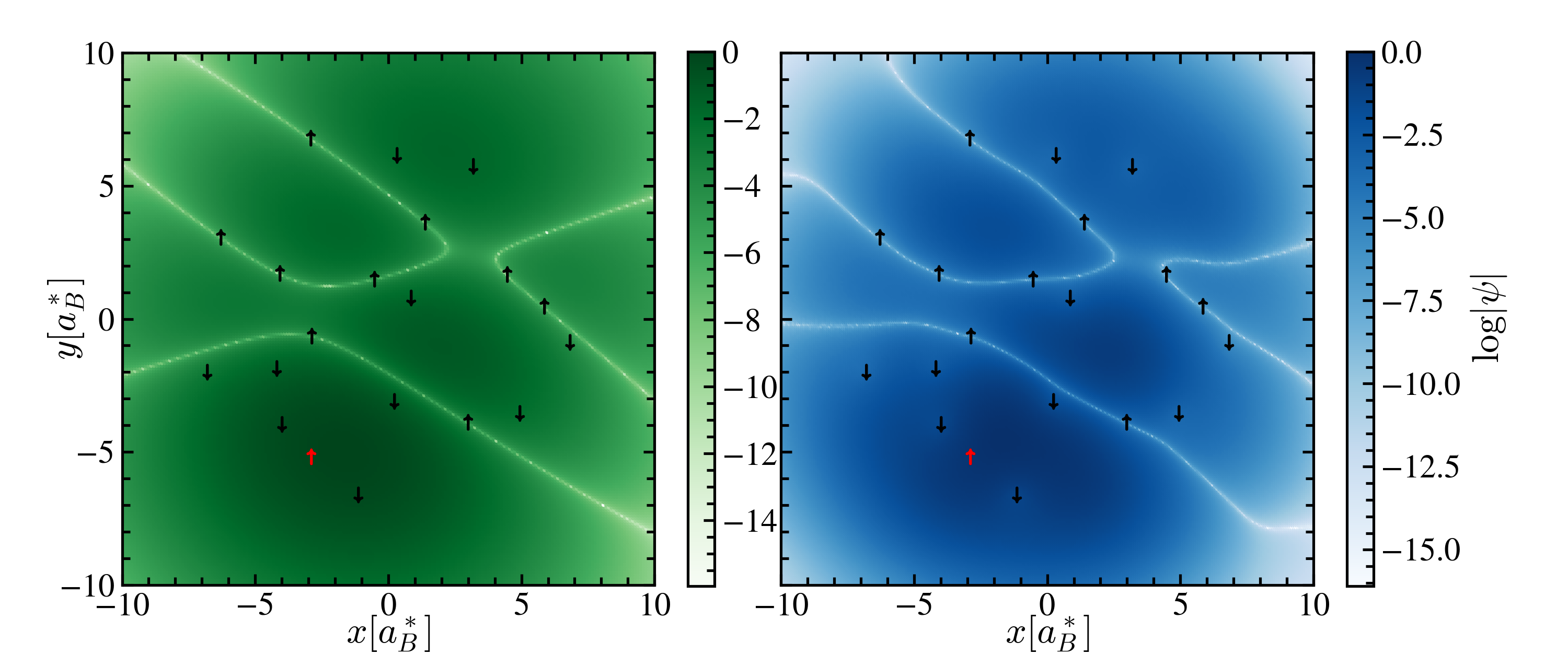}}
\subfloat[]{\includegraphics[width=0.5\linewidth]{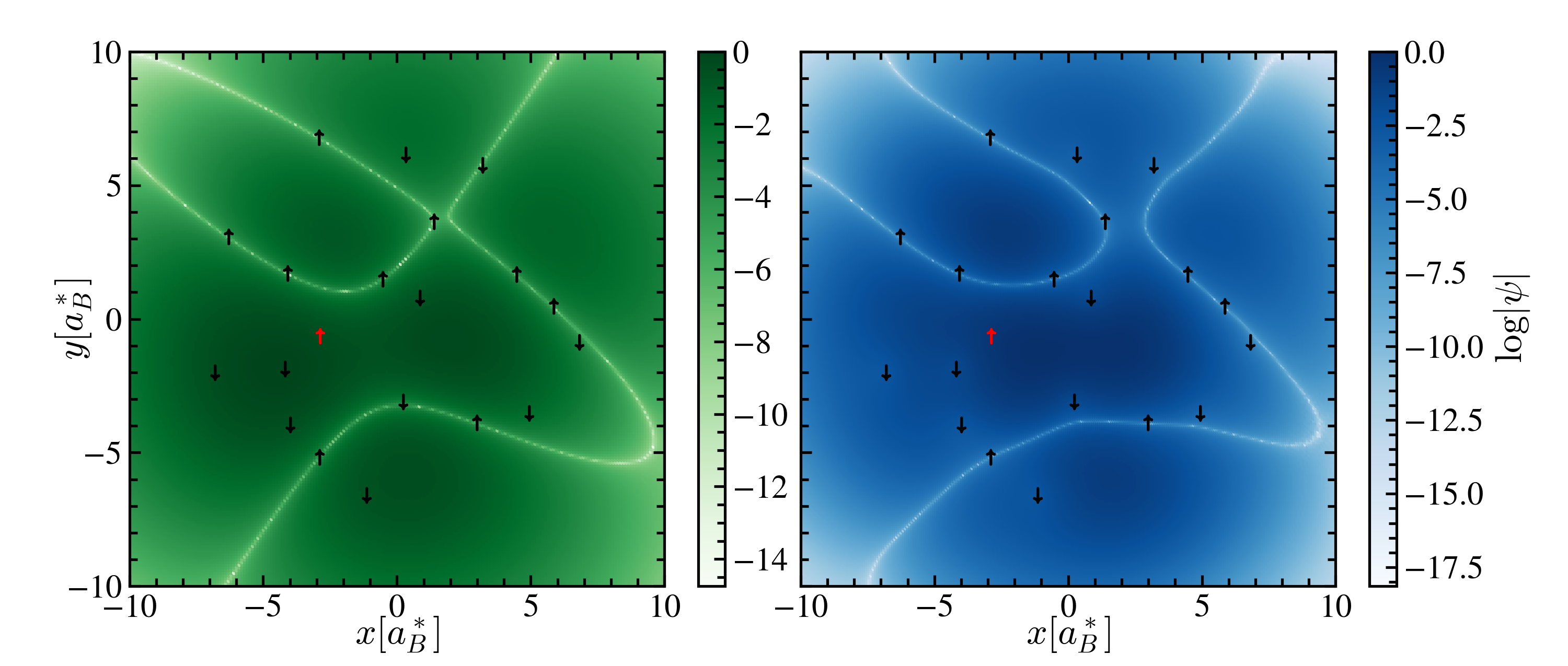}}

\subfloat[]{\includegraphics[width=0.5\linewidth]{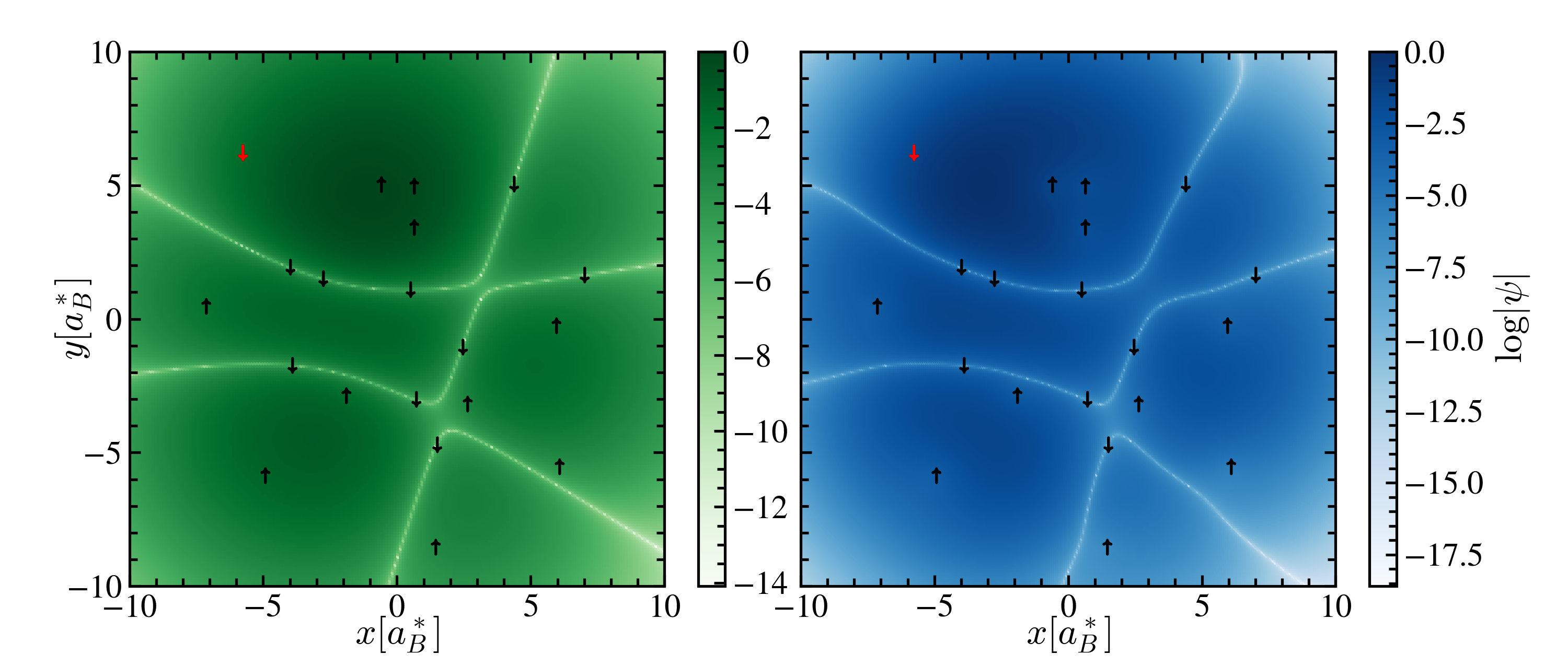}}
\subfloat[]{\includegraphics[width=0.5\linewidth]{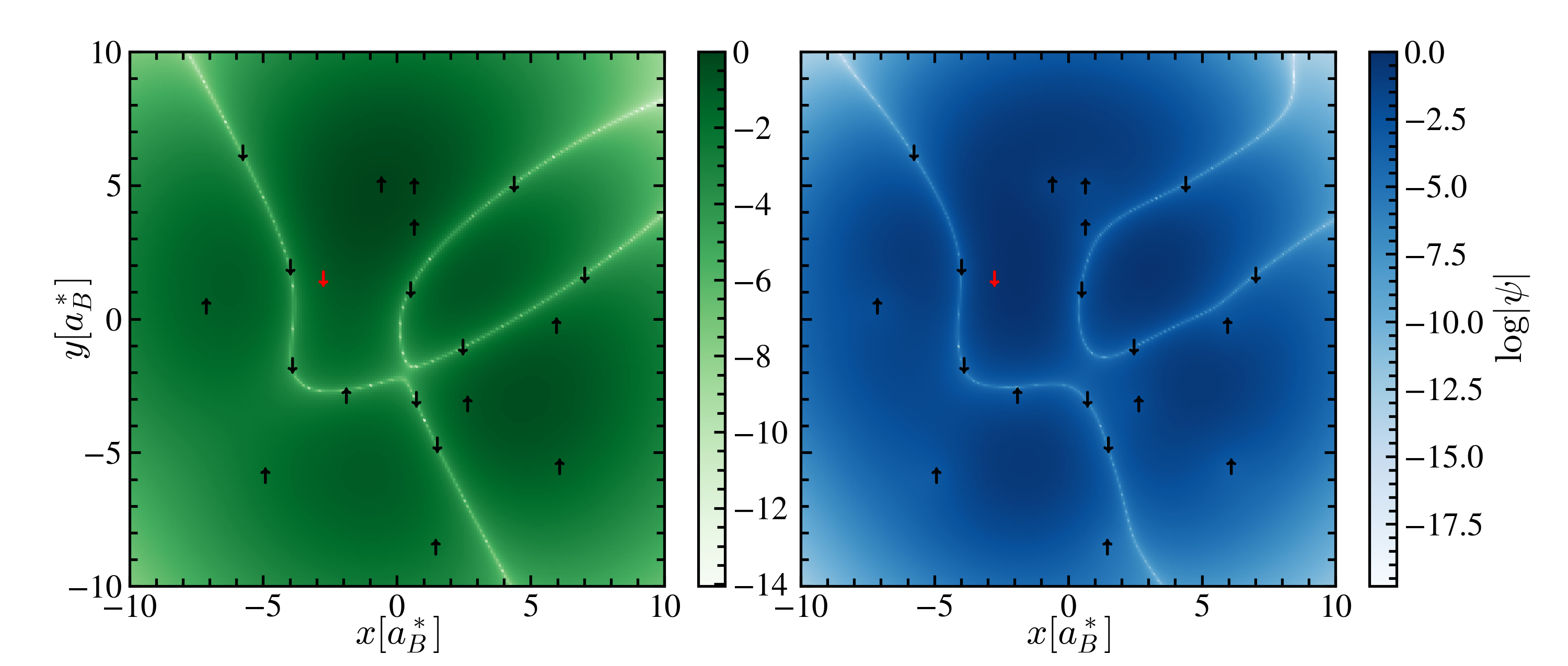}}
\caption{(a-d) Nodal structures of a harmonic oscillator trial function (left) compared to those of the neural network representation of the machine-learned wavefunction used in this study (right). In (d), a different configuration from the previous ones is considered. Each subfigure corresponds to a different spatial point used for analysis.}
\label{fig:nodal}
\end{figure*}

The nodal structure of a trial function of the harmonic oscillator form is compared with that of the neural network representation of the machine-learned wavefunction used in this study. \fig{nodal}
presents four two-dimensional color maps of the hyperplane of our optimized trial function for a QD with 20 electrons, with white lines indicating the nodal hypersurfaces in each subfigure. The figure was constructed by selecting an equilibrated configuration from our trial function and fixing all electron coordinates except one (red arrow), which was then moved over the QD. The wave function values are shown in the color scale, with the numerical axis representing the coordinates of the moving electron. 

For a direct comparison between the neural network-generated wavefunction and the traditional trial function, we used the same initial electron configuration in both cases. This ensures that any observed differences in the nodal structure arise solely from the distinct functional forms of the two wavefunctions, rather than variations in initial conditions. In the case of the traditional trial function, the optimal variational parameter—which minimizes the total energy within the constraints of the chosen ansatz—was determined through an independent optimization process before being applied to the comparison. This approach allows for a fair assessment of how well each method captures the true nodal topology and highlights the advantages of the neural network representation in refining wavefunction accuracy.

%

\end{document}